\documentclass[twocolumn]{aastex631}

\begin{document}

\title{Magnetic accretion flow explains the hysteresis q-diagram seen in outbursts of black hole low-mass X-ray
binaries}

\author[0000-0001-8674-2336]{Jiahui Huang}
\affiliation{Center for Computational Sciences, University of Tsukuba, 1-1-1 Ten-nodai, Tsukuba, Ibaraki 305-8577, Japan}

\author[0000-0001-7584-6236]{Hua Feng}
\email{hfeng@ihep.ac.cn}
\affiliation{Key Laboratory of Particle Astrophysics, Institute of High Energy Physics, Chinese Academy of Sciences, Beijing 100049, China}

\author[0000-0002-2705-4338]{Lian Tao}
\email{taolian@ihep.ac.cn}
\affiliation{Key Laboratory of Particle Astrophysics, Institute of High Energy Physics, Chinese Academy of Sciences, Beijing 100049, China}



\begin{abstract}

Black hole low-mass X-ray binaries undergo quiescence-outburst cycles. During the outbursts, they typically go through a q-shaped pattern in the hardness-intensity diagram (HID), known as the hysteresis q-diagram, while the physical nature is still unknown. We argue that the hysteresis q-diagram can be well explained with a recently proposed magnetized accretion disk model. The model takes into account the saturated magnetic pressure and predicts that the standard Shakura-Sunyaev disk (SSD) has an inner truncation at relatively low accretion rates, filled with an optically thin, hot accretion flow that resembles the advection-dominated accretion flow (ADAF) inside. Given a perturbation of accretion rate, the variation of the truncation radius can be derived as a result of thermal equilibrium by comparing the heating and cooling rates. We show that the truncation radius displays a hysteresis effect in response to the variation of mass accretion rate. As a result, the spectral hardness due to competition of the soft SSD and hard ADAF components is also hysteresis along with the rise and decay of the mass accretion rate or source intensity, leading to a q-shaped diagram in the HID.

\end{abstract}



\section{Introduction} \label{sec:intro}

Black hole low-mass X-ray binaries (BH-LMXBs) are powered by accretion onto stellar mass black holes from a stellar companion \citep{Frank2002}. 
They mostly stay in the quiescent state with very low luminosities, but some of them may exhibit major outbursts, during which the luminosity can increase by orders of magnitude. 
In the outburst, BH-LMXBs display complex spectral and timing behaviors in the X-ray band, with correlated multi-wavelength properties, classified into different emission states \citep{Fender2004, Homan2005, Remillard2006, Done2007, Corral-Santana2016}. 

Despite complicated state transitions in the outburst, their spectral evolution may trace a common pattern in the X-ray hardness-intensity diagram \citep[HID;][]{Homan2005}. 
If one extracts the source count rate in the soft and hard bands, respectively, one may calculate the hard-to-soft flux ratio (hardness) and determine the source status in the HID along with the source count rate (intensity). 
The boundary between the soft and hard bands can be chosen around a few keV \citep{Homan2005}, such that the hard flux represents emission from the hot corona, while the soft band contains a significant fraction of flux from the cool accretion disk. 
In major outbursts, BH-LMXBs typically follow a q-shaped pattern in the HID, also referred to as the hysteresis q-diagram, e.g., seen in GX~339--4 \citep{Zdziarski2004}, GRS~1915+105 \citep{Fender2004}, MAXI~J1820+070 \citep{Kara2019}, and MAXI~J1348--630 \citep{Tominaga2020}. 
This suggests that the spectral properties of BH-LMXBs are not a monotonic function of the mass accretion rate, and the soft to hard transition usually happens at a much lower luminosity than the hard to soft transition \citep{Maccarone2003}. 

To the first order of approximation, the emission state of BH-LMXBs can be understood as competition between two emission components: a soft thermal component from an optically thick, geometrically thin accretion disk, or the so-called Shakura-Sunyaev disk \citep[SSD;][]{Shakura1973}, and a hard Comptonization component from the corona or advection-dominated accretion flow \citep[ADAF;][]{Narayan1994, Narayan1995}. 
In the hard state, many observations have revealed a disk inner radius larger than the innermost stable circular orbit (ISCO) predicted by the theory of general relativity, suggesting that the SSD may have an inner truncation \citep{Esin2001, Zdziarski2004, Tomsick2009, Garcia2015}. 
Some models for disk evaporation may explain the inner disk truncation \citep{Liu2006, Meyer2007, Qiao2010}. 
A manually connected inner ADAF plus outer SSD model was proposed to reconcile with the observations \citep{Esin1997}.
However, simply combining the ADAF and SSD cannot naturally explain the hysteresis effect seen in the HID. 
There is at least another parameter besides the accretion rate responsible for the hysteresis effect \citep{Homan2001}. 


In this paper, we argue that a recently proposed accretion disk model that takes into account the magnetic pressure \citep{Huang2023_model} can naturally explain the hysteresis effect seen in the outburst of BH-LMXBs.

\section{Model}
\label{sec:model}

Here we briefly review the model of \citet{Huang2023_model}. 
Numerical simulations reveal that the magnetic pressure is the dominant component in supporting the disk vertically when the mass accretion rate is in the range of 0.01--1 times the critical mass accretion rate, which is the rate just needed to power the Eddington limit \citep{Lancova2019, Jiang2019, Huang2023_sim}. 
In the accretion process, the toroidal magnetic fields will be amplified due to dynamo of the magneto-rotational instability \citep[MRI;][]{Balbus1991, Balbus1998}, and saturates at a pressure, $P_{\rm B}=\rho V_{\rm K} c_{\rm g}$, when the Alfv\'{e}n velocity $V_{\rm A}$ approaches $\sqrt{V_{\rm K} c_{\rm g}}$ \citep{Pessah2005}, where $\rho$ is the gas density, $V_{\rm K}$ is the local Keplerian velocity, and $c_{\rm g}$ is the gas sound speed.
The saturated magnetic pressure gives a reasonable fit to the simulation results \citep{Huang2023_sim}. 


 
Inspired by the simulation results, \citet{Huang2023_model} proposed a new accretion disk model that takes into account the saturated magnetic pressure in addition to the gas and radiation pressures, and also outflows. 
In the simulation, the outflows are driven by the magnetic and radiation (in the case of high accretion rate) pressure gradients \citep{Huang2023_sim}. 
In the model, the accretion rate is assumed to have a power-law dependence with radius to account for the mass and energy loss due to outflows, for simplicity. 
The model assumes that the mass loss due to outflows is higher in the ADAF than in the SSD, which is consistent with both simulations \citep{Ohsuga2011, Ohsuga2014, Yuan2015} and observations \citep{Tombesi2010, Tombesi2014, Munos2019}. 
The model neglects the energy loss due to Poynting flux because no large-scale magnetic fields or general relativity is implemented, which is a caveat of this simplified 1D model.
The model yields a solution that the inner SSD can no longer reach thermal equilibrium at relatively low accretion rates due to insufficient radiative cooling. 
Instead, an optically thin, hot accretion solution is found at small radii, which resembles the ADAF solution. For convenience, we also call this solution an ``ADAF'', although the wind cooling is the dominant cooling mechanism in this optically thin flow instead of the advective cooling.
Such an outer SSD and inner ADAF configuration is no longer an assumption, but a result of global solutions of the model, and can naturally explain the observed disk truncation mentioned above.

\begin{figure}
\centering
\includegraphics[width=0.9\columnwidth]
{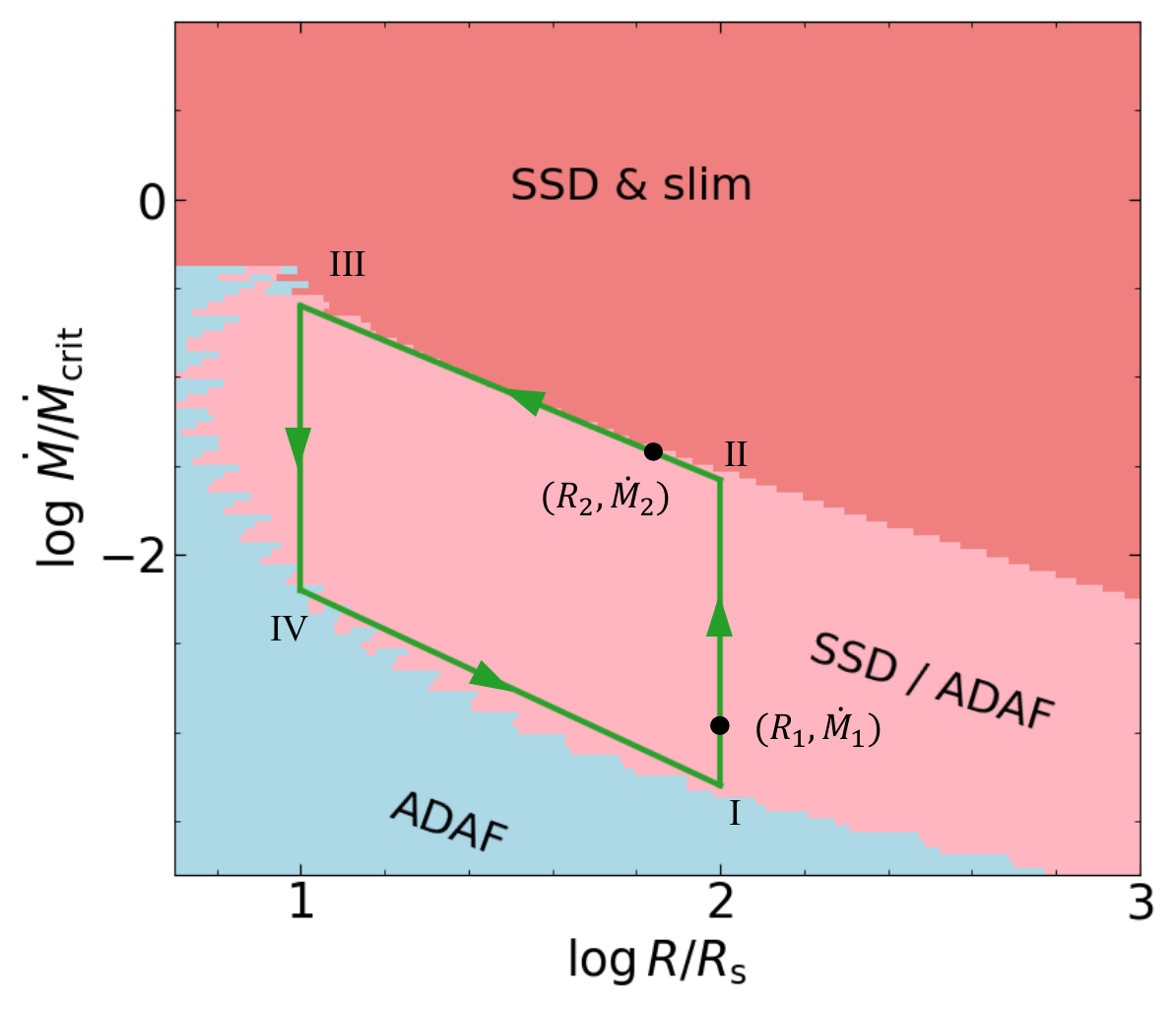}
\includegraphics[width=0.9\columnwidth]
{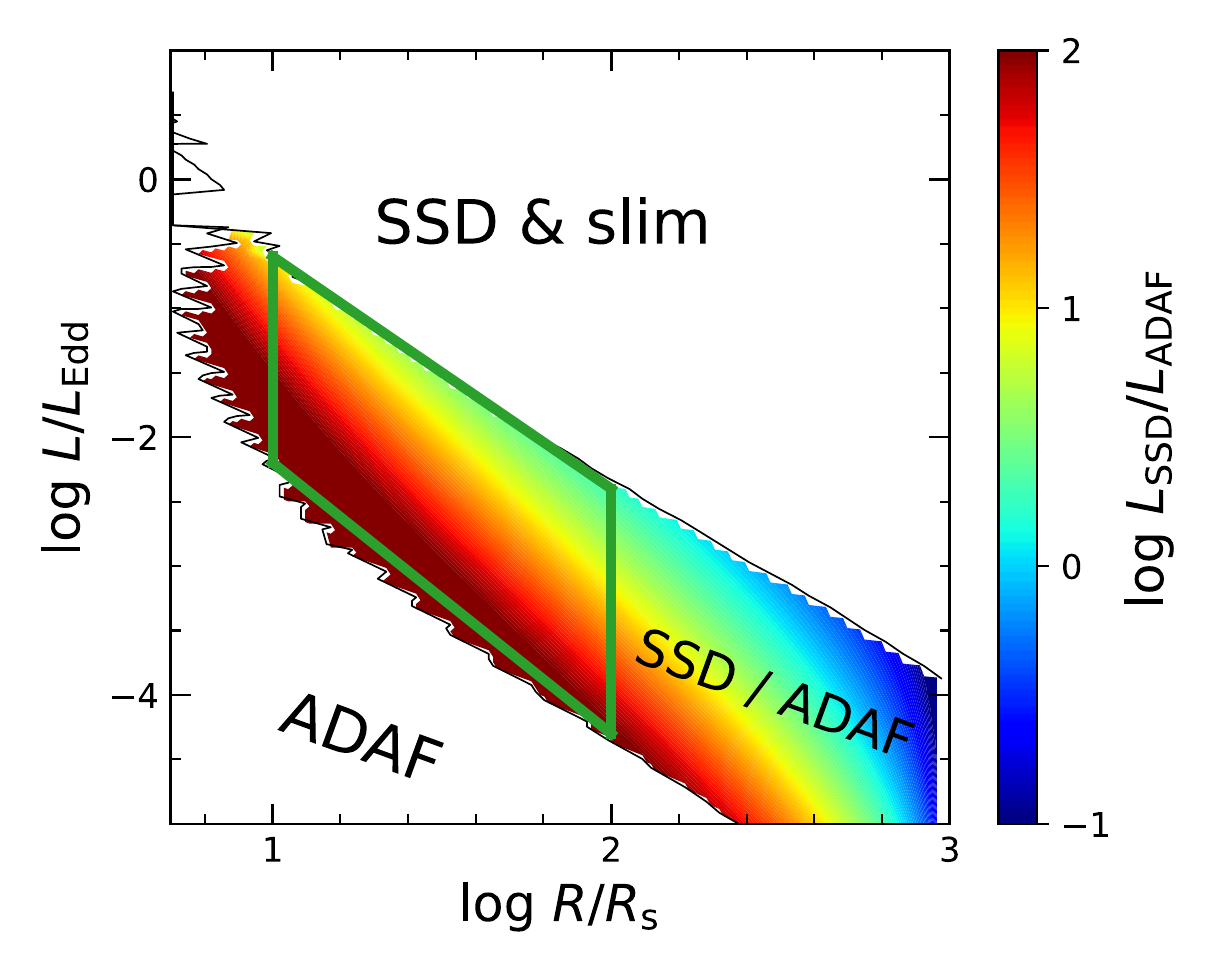}
\caption{Solutions of the magnetic disk model as a function of radius and mass accretion rate (top) or luminosity (bottom), adapted from Figure~6 in \citet{Huang2023_model}. 
$L_{\rm Edd}$ is the Eddington luminosity.
$\dot{M}_{\rm crit} = L_{\rm Edd} / 16c^2$ is the critical mass accretion rate to power $L_{\rm Edd}$. 
$R_{\rm s} = 2GM_{\rm BH}/c^2$ is the Schwarzschild radius. In the top panel, the red area marks the SSD or slim disk solution, where radiative cooling dominates; the blue area marks the ADAF solution; in the pink area, either the SSD or ADAF solution can exist. 
The green arrows show an example how the truncation radius varies along with the increase (I $\to$ II $\to$ III) and decrease (III $\to$ IV $\to$ I) of mass accretion rate or luminosity during an outburst. $(R_1, \dot{M}_1)$ and $(R_2, \dot{M}_2)$ are two arbitrary locations along the track used for perturbation analysis. In the bottom panel, the colormap shows the ratio of the SSD luminosity to ADAF luminosity in the region where the two solutions coexist.
\label{fig:route}}
\end{figure}

Given a mass accretion rate, the model predicts three types of solutions at each radius, reminiscent of the canonical SSD, ADAF, and slim disk solutions \citep{Shakura1973, Abramowicz1988, Narayan1994, Narayan1995}, respectively. 
A solution diagram as a function of mass accretion rate and radius is shown in Figure~\ref{fig:route}, in which the pink area labeled with ``SSD/ADAF'' indicates that either solution may exist in this parameter space. For continuity, the connection radius between the outer SSD and inner ADAF, i.e., the truncation radius, must fall into this region. Such an alternative solution is confirmed by recent numerical simulations of a magnetically arrested disk \citep{Scepi2024}. The corresponding solution diagram as a function of luminosity and radius is also shown in Figure~\ref{fig:route}.

\section{Evolution of truncation radius} 
\label{sec:truncation_radius}

In the very beginning of the outburst when the accretion rate is low, the whole accretion flow is in an ADAF mode. With the increase of mass accretion rate, the innermost region remains in ADAF while the outermost region becomes SSD. This suggests that the truncation radius falls into the SSD/ADAF (pink) region in Figure~\ref{fig:route}. Supposing that the accretion flow starts to evolve from an arbitrary location I in Figure~\ref{fig:route}, we argue that, along with the variation of mass accretion rate in a major outburst, the truncation radius shows a hysteresis effect and may follow a closed cycle in the diagram indicated by the green arrows (I $\to$ II $\to$ III $\to$ IV $\to$ I) in Figure~\ref{fig:route}.

\begin{figure}
\centering
\includegraphics[width=0.75\columnwidth]{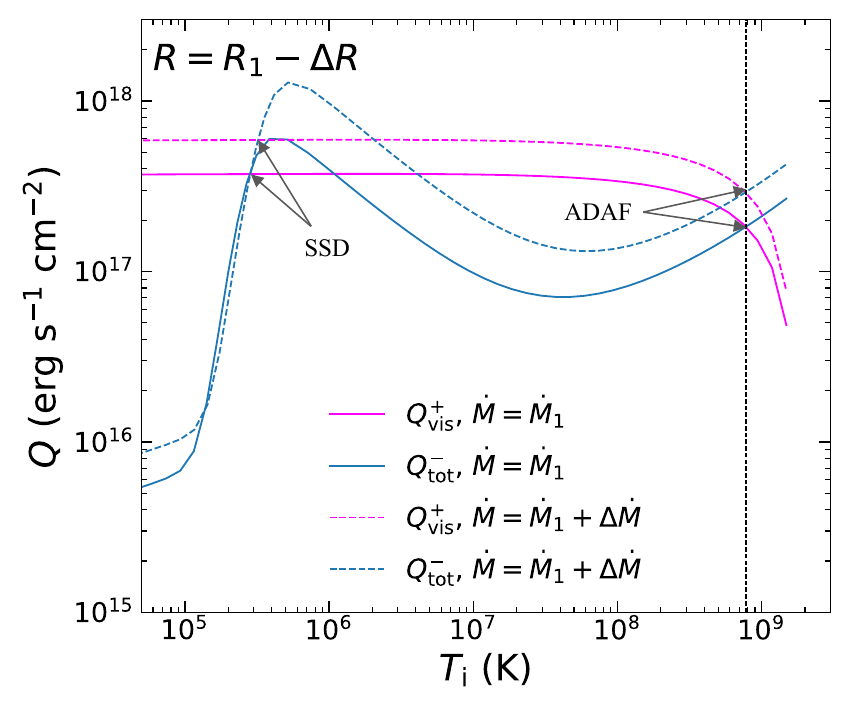}
\includegraphics[width=0.75\columnwidth]{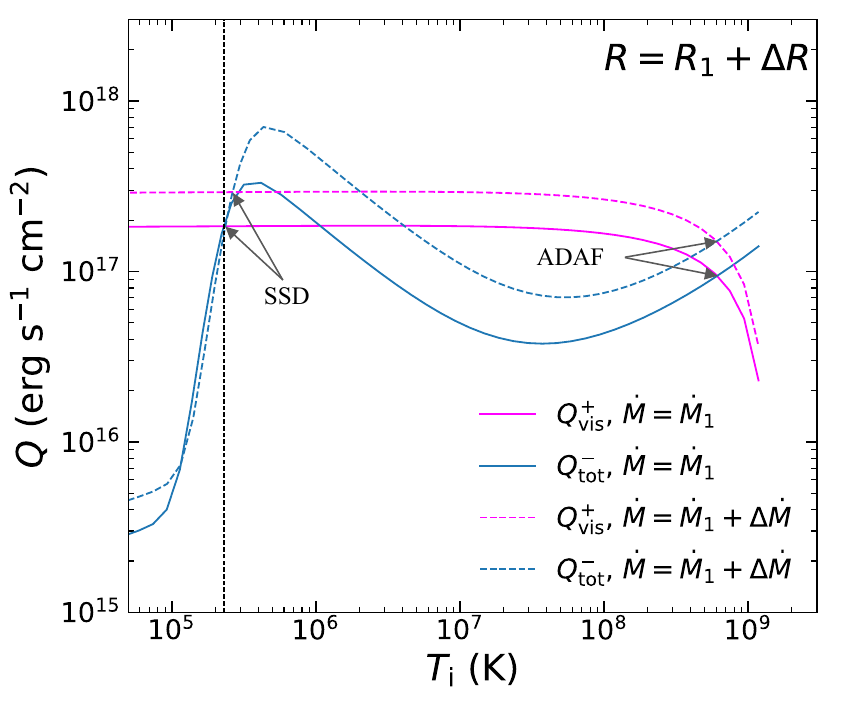}
\caption{Total heating and cooling rates as a function of ion temperature at $R_1 \pm \Delta R$ around the location $(R_1, \dot{M}_1)$ in Figure~\ref{fig:route}. The magenta and blue solid lines are the original total heating rate $Q_{\rm vis}^+$ and cooling rate $Q_{\rm tot}^-$, respectively. The magenta and blue dashed lines are the total heating rate and cooling rate, respectively, after perturbation. The black dashed vertical line shows the original disk temperature. After perturbation, the disk solution at the given radius remains unchanged.
\label{fig:perturbation_1to2}}
\end{figure}

\textbf{A) I--II.} 
Given an initial truncation radius $R_1$ and mass accretion rate $\dot{M}_1$ along the track I--II (marked in Figure~\ref{fig:route}), we assume a perturbation of mass accretion rate to $\dot{M}_1 + \Delta \dot{M}$, and examine the total heating and cooling curves as a function of ion temperature at two radii, $R_1 \pm \Delta R$, around the truncation radius, shown in Figure~\ref{fig:perturbation_1to2}. 
At each radius, there are three interception points between the heating and cooling curves, representing three thermal equilibrium solutions, i.e., the SSD, Shapiro-Lightman-Eardley \citep{Shapiro1976}, and ADAF solutions, respectively from left to right \citep{Huang2023_model}. 
At $R_1 - \Delta R$, the disk is in the form of ADAF before perturbation and the rightmost intersection point is the valid solution (marked by a vertical dashed line). 
After the perturbation, the disk is still at thermal equilibrium ($Q_{\rm vis}^+ = Q_{\rm tot}^-$) at the similar ion temperature, suggesting that the disk at $R_1 - \Delta R$ will remain in ADAF. 
At $R_1 + \Delta R$, the disk is in the form of SSD before perturbation, corresponding to the leftmost interception point, also marked by a vertical dashed line. 
After the perturbation, the increase of heating rate is higher than the increase of cooling rate, resulting in a higher ion temperature for thermal equilibrium, but the disk is still in SSD. 
Thus, the truncation radius will remain constant (along I--II) in Figure~\ref{fig:route} in response to variations of mass accretion rate. 


\begin{figure}
\centering
\includegraphics[width=0.75\columnwidth]{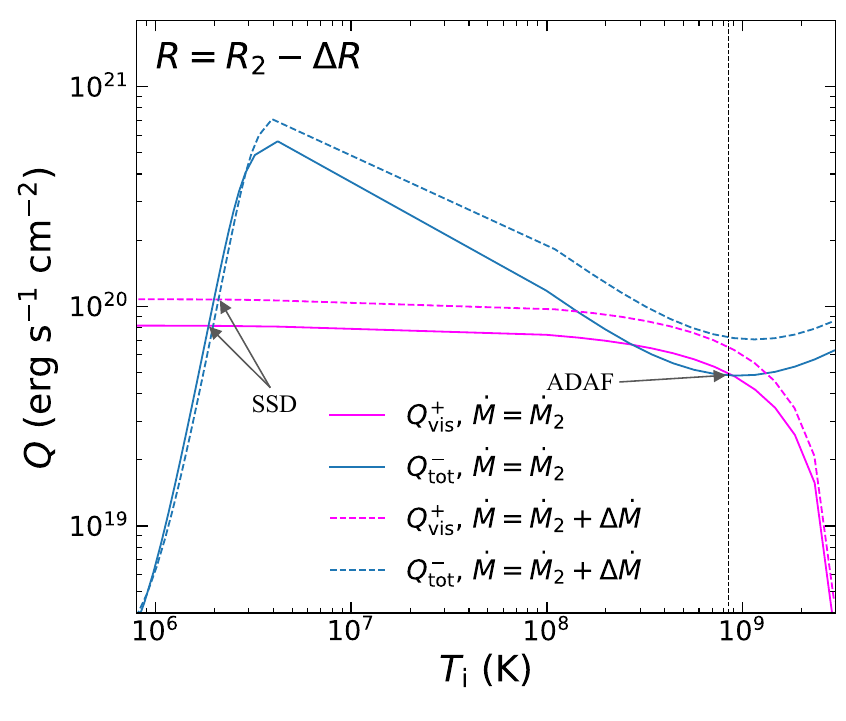}
\includegraphics[width=0.75\columnwidth]{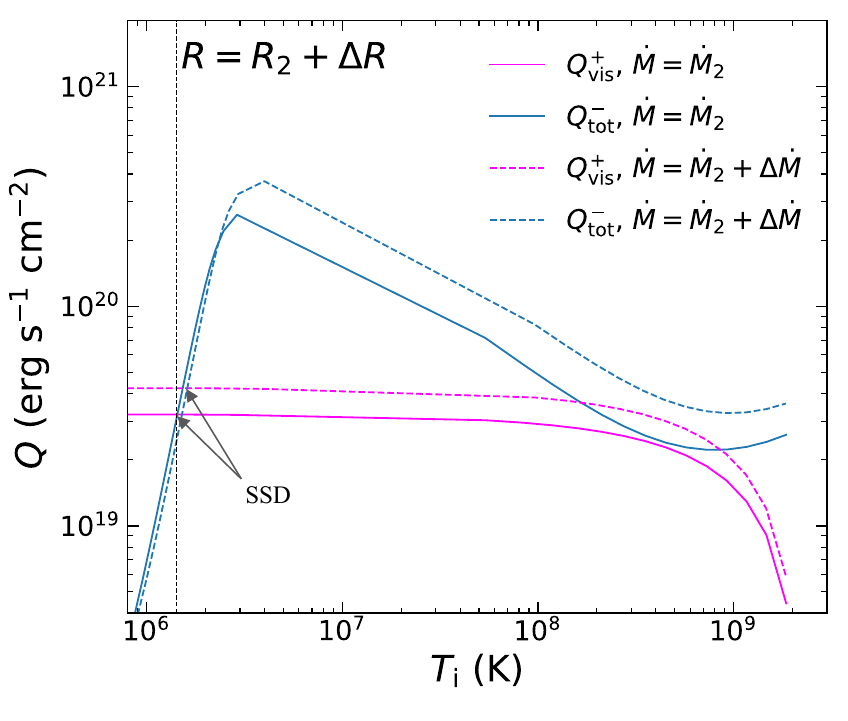}
\caption{Total heating and cooling rates as a function of ion temperature at $R_2 \pm \Delta R$ around the location $(R_2, \dot{M}_2)$ in Figure~\ref{fig:route}. The lines have the same meaning as in Figure~\ref{fig:perturbation_1to2}. The disk solution changes from ADAF to SSD after perturbation at $R_2 - \Delta R$, but remains in SSD at $R_2 + \Delta R$.
\label{fig:perturbation_2to3}}
\end{figure}

\textbf{B) II--III.} 
When the mass accretion rate is high enough and the system moves to location II in Figure~\ref{fig:route}, i.e., the upper boundary of the ADAF/SSD region, a further tiny increase of mass accretion rate will make the truncation radius move inward along the track II-III. We demonstrate this as follows. We assume an arbitrary accretion rate $\dot{M}_2$ with a tiny increase to $\dot{M}_2 + \Delta \dot{M}$ (marked in Figure~\ref{fig:route}). 
Similar to the discussions above, we also plot the heating and cooling curves at two radii just inside and outside the truncation radius $R_2$ (shown in Figure~\ref{fig:perturbation_2to3}). 
At $R_2 - \Delta R$, the increase of cooling rate exceeds the increase of heating rate after perturbation. 
However, the only thermal equilibrium solution becomes the leftmost interception point, corresponding to the SSD solution. 
The ADAF to the left the truncation radius can no longer sustain a thermal equilibrium, and the extra radiative cooling over viscous heating causes the hot flow to condense into an SSD. 
Thus, the truncation radius cannot move into the red zone. If the truncation radius evolves into the pink area, then similar to the case of I--II, a further increase of the accretion rate will push it upward and eventually reach the upper boundary. 
Based on the discussion above, the truncation radius can only evolve along the upper boundary of the ADAF/SSD region, shown as the track II--III in Figure \ref{fig:route}.


\begin{figure}
\centering
\includegraphics[width=0.75\columnwidth]{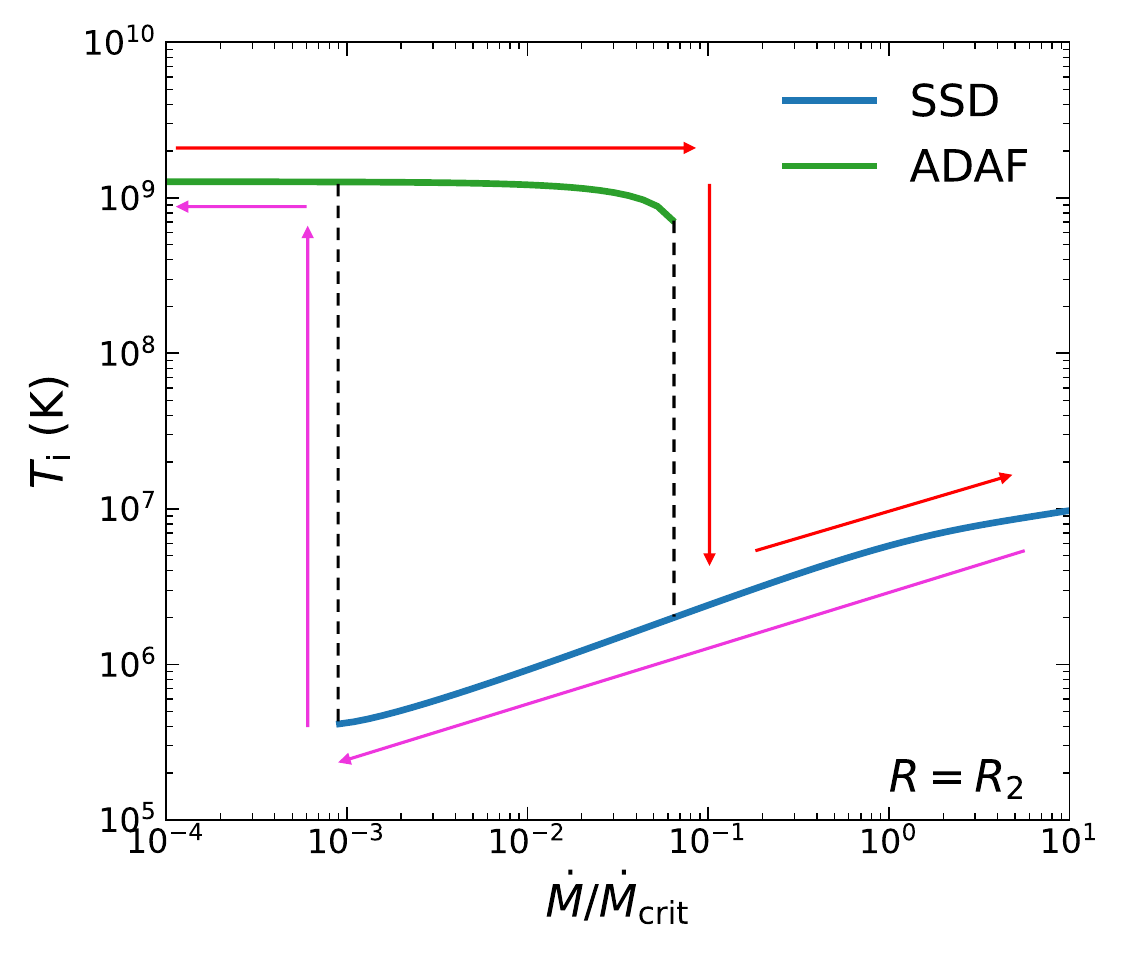}
\caption{Ion temperature of the accretion flow as a function of accretion rate at an arbitrary radius. The green line represents the ADAF branch, while the blue line represents the SSD branch. The red (magenta) arrows show how the disk temperature varies when the accretion rate increases (decreases).
\label{fig:T_vs_mdot}}
\end{figure}

\textbf{C) III---IV.}
If the mass accretion rate starts to decrease, e.g., at location III in Figure~\ref{fig:route}, the same argument for the track I--II works here, and the truncation radius remains constant with decreasing accretion rate (along III--IV). 

\textbf{D) IV--I.}
Similarly, a further decrease of mass accretion rate at the location IV will cause the truncation radius to move outward along the lower boundary of the ADAF/SSD region (along IV--I), as the truncation cannot occur in the ADAF region.  

To further demonstrate the hysteresis effect, we plotted the ion temperature of the accretion flow at an arbitrary radius. 
When the accretion rate is low at the beginning of the outburst, only the ADAF solution exists. 
As the accretion rate increases, the disk remains in ADAF until at some point with a sudden transition to SSD. 
Then, the disk can go back to ADAF only at an even lower mass accretion rate.
This is consistent with the disk model described in \citet{Scepi2024}.

The location III represent the highest mass accretion rate during the outburst. A higher maximum mass accretion rate leads to a smaller truncation radius toward the ISCO radius. For a failed outburst, the maximum mass accretion rate is not high enough, such that the SSD cannot grow all the way to ISCO. This has been observed in the BH-LMXB GX~339--4, which shows a large truncation radius during the decay of a failed outburst \citep{Wang2018}.

\section{formation of q-diagram}
\label{sec:q-diagram}


It is straightforward to relate the truncation radius with spectral hardness. As mentioned above, soft X-rays originate from the SSD while hard X-rays are from the ADAF. An increasing truncation radius leads to a strengthened ADAF and weakened SSD and subsequently a harder spectrum, and vice versa. Therefore, the I--II track in Figure~\ref{fig:route} represents a hard spectrum and III--IV corresponds to a soft spectrum. The source intensity is determined by the mass accretion rate. Therefore, the evolution of a BH-LMXB during an outburst in the HID can be summarized as follows. In the rising phase, the source enters the outburst via the low-hard state (I) and goes for the high-hard state (II), and then transitions to the high soft state (III).  
In the decay, the source will remain in the soft state until a relatively low luminosity (IV), before transition back to the low-hard state (I). 

\begin{figure}
\centering
\includegraphics[width=0.75\columnwidth]{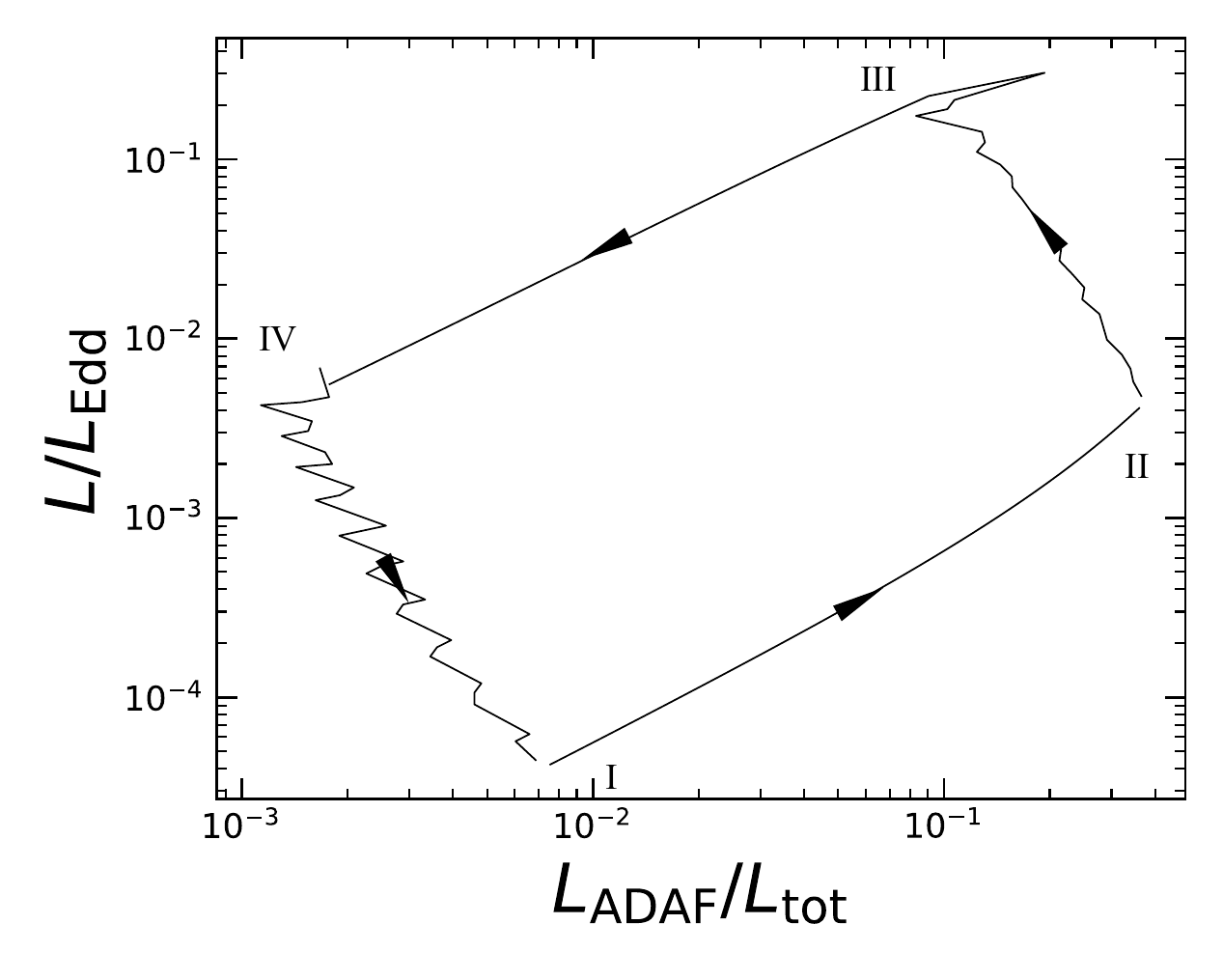}
\caption{Disk fraction luminosity diagram (DFLD) along the track I-II-III-IV-I.
\label{fig:DFLD}}
\end{figure}

We calculated the Eddington ratio vs.\ the fractional luminosity of the ADAF component, depicted in Figure \ref{fig:DFLD} known as the disk fraction luminosity diagram (DFLD), which shows a full hysteresis cycle. 
Here, only the bremsstrahlung radiation in the ADAF is considered \citep{Narayan1995,Huang2023_model}. 
The ADAF luminosity fraction, to some extent, may approximate the spectral hardness, but detailed spectral modeling is still needed for an accurate calculation. 
For example, at state III, the high luminosity fraction of ADAF is associated with an optical depth exceeding unity, suggesting that the accretion flow at this point may resemble a slim disk solution and primarily produce soft photons \citep[see Equation 16 in][]{Huang2023_model}. 
At state I, although the ADAF luminosity fraction is low, a large truncation radius may have resulted in a large filling factor for soft photons from the outer or possibly underlying SSD, producing a hard spectrum. 
The absence of detailed modeling of the emergent spectrum is a limitation of this work and is reserved for future improvement of this model.

\section{Discussion}

In this paper, we demonstrate that the recently proposed magnetic accretion disk model \citep{Huang2023_model} can naturally explain the hysteresis q-diagram seen in the outbursts of BH-LMXBs. The key or basis for the model to work is that it can predict an accretion flow comprising an inner ADAF plus an outer SSD from a global solution.

Observations show that the hysteresis effect appears only in the HID, but not in the hardness-rms diagram \citep{Belloni2009, Motta2011}, where the fractional rms variability is correlated with hardness. As the variability mainly originates from the ADAF component, the relative contribution of ADAF/SSD determines the fractional rms variability. Therefore, the source at the top branch (hard-to-soft transition) and the bottom branch (soft-to-hard transition) in the q-diagram shows similar fractional rms variability if the truncation radius is the same, leading to a simple correlation without hysteresis effect.

The critical luminosity for state transition is a useful for comparison with observations. As shown in Figure \ref{fig:DFLD}, the accretion flow is dominated by the hard ADAF at $L<0.1L_{\rm Edd}$. However, when the luminosity approaches $0.1L_{\rm Edd}$, the optical depth approaches unity and the ADAF solution will transition to the slim disk solution, leading to dominant soft emission. Thus, the hard-to-soft transition may occur in the II-III track at a luminosity range $\gtrsim 0.1L_{\rm Edd}$.
This is in general consistent with observations of BH-LMXBs \citep[see Figure 14 in][]{Tetarenko2016}. The hard-to-soft state transitions of BH-LMXBs may have different dependence on flux. The source count rate may increase, e.g. XTE~1550-564 \citep{Dunn2010}, keep constant, e.g. GX~339-4 in the 2002/2003 outburst \citep{Zdziarski2004}, XTE~J1908-094 in the 2002/2003 outburst \citep{Gogus2004}, or decrease, e.g. XTE~J1650-500 in the 2001/2002 outburst \citep{Homan2005} during the hard-to-soft state transition. Our model can naturally explain the increasing flux and marginally constant flux during the hard-to-soft transition, but has difficulties in explaining the decreasing type.

Our model also implies that, during the decay in the soft thermal state, one expects a constant inner disk radius, which, however, is not necessarily the ISCO radius, if the maximum accretion rate is not high enough to push the SSD all the way into ISCO. If this is true, estimation of black hole spin based on the assumption that the disk extends all the way to ISCO in the thermal state will be problematic. We mention that this could be a caveat for a simplified 1D model, because the self-similar assumption in \citet{Huang2023_model} is no longer valid at the innermost region of the accretion disk. 
Also, GX~339-4 showed a decreasing cutoff energy (or corona temperature) in the hard state along with increasing mass accretion rate \citep{Motta2009}. 
However, our model predicts an almost constant temperature of the ADAF component along the I-II track, which could be another weakness of the model that does not take into account the radiative transfer that may lead to corona cooling with increasing disk flux.

In the literature, the jet emitting disk-standard accretion disk (JED-SAD) model \citep{Ferreira1997, Petrucci2010, Marcel2018, Marcel2018a, Marcel2019, Jacquemin2019, Marcel2022, Barnier2022} gives similar results and can also explain the hysteresis cycle of X-ray binary outbursts. 
Compared with ours, their model shows similar global solutions, and also similar optically thin to optically thick transitions at a specific disk radius along with variations of the mass accretion rate \citep[see Figure 11 in][]{Marcel2018}. 
The JED-SAD model assumes strong magnetic fields and treat them directly in the Maxwell equations, while our model simply considers the magnetic pressure originated from MRI turbulence. 
The similar conclusions may be a result of a similar magnetization degree in both models.
Our model predicts a truncation of SSD \citep{Huang2023_model}, which, however, is not present in the JED-SAD model.

Recent observations with the Imaging X-ray Polarimetry Explorer show that the emission of X-ray binaries in both the hard and soft states can be significantly polarized. This seems to suggest the absence of strong coherent large-scale magnetic fields ($B>5\times10^6$~G) that will cause depolarization due to the Faraday rotation effect \citep{Barnier2024}. However, in our model, the strong magnetic pressure is associated with the turbulent magnetic fields generated by MRI \citep{Huang2023_sim}, in which the Faraday depolarization may be largely suppressed compared with in ordered fields.

To conclude, we find that the magnetic accretion disk model provides a reasonable explanation for the hysteresis q-diagram to the first order of approximation. 

\begin{acknowledgments}

We thank the anonymous referee for insightful comments. We acknowledge funding support from the National Natural Science Foundation of China under grant Nos. 12025301 and 12122306, and the Strategic Priority Research Program of the Chinese Academy of Sciences.

\end{acknowledgments}

\bibliography{sample631}{}
\bibliographystyle{aasjournal}

\end{document}